\documentclass{article}
\usepackage[utf8]{inputenc}

\usepackage{url}

\usepackage{natbib}
\bibpunct{(}{)}{;}{a}{}{,}%
\usepackage{graphicx}
\usepackage{amsmath,amssymb}
\usepackage{multirow}
\usepackage{longtable}

\usepackage{epsfig}
\usepackage{setspace} 

\usepackage[b5paper]{geometry}
\usepackage{authblk}
\title{Superluminous quasars and mesolensing}
\author[1,2]{Alexander Raikov}
\author[3]{Nikita Lovyagin}
\author[4]{Vladimir Yershov}
\affil[1]{Special Astrophysical Observatory, Russian Academy of Sciences\linebreak Niznii Arkhyz, 369167 Russia}
\affil[2]{Pulkovo Observatory\linebreak 65(1) Pulkovskoye shosse, St. Petersburg, 196140, Russia}
\affil[3]{Saint Petersburg State University\linebreak 7-9 Universitetskaya emb., St. Petersburg, 199034, Russia}
\affil[4]{Mullard Space Scince Laboratory,\linebreak Holmbury St.Mary, RH5 6NT, United Kingdom}
\date{}

\begin{document}
  \maketitle

\begin{abstract}
Observed magnitudes  of many quasars with redshifts exceeding 
$z=5$ correspond to luminosities $L_{\rm bol} > 10^{14}\,L_\odot$. 
The standard mechanism of quasar energy release by accretion suggests that 
masses of superluminous quasars should  
exceed $10^{10}\,M_\odot$. On the other hand, the age of these objects 
in the standard cosmological model is below one billion years, which 
is too short to explain their formation in the early Universe.  
Many quasars are known to be gravitationally lensed; 
showing multiple images of the same object. In the case of remote 
quasars with no multiple images, it is still possible 
that they are also gravitationally lensed by foreground objects of intermediate 
masses, such as globular clusters or dwarf  galaxies. Such mesolensing 
would result in essential amplification of quasar brightnesses, 
subject to geometrical configuration between the lens and the lensed object. 
Here we estimate the fraction of quasars 
whose brightness might have been amplified by gravitational lensing. 

{\bf Keywords: } quasars, luminosity function, gravitational lensing

{\footnotesize To be published in Proc. conf. VAK-2021, August 23-28, 2021, Sternberg Astronomical Institute, Moscow

}
\end{abstract}

%
Since  \cite{arp87}, it has been known that there is a statistically significant 
over-density of quasars within areas of the sky near foreground galaxies. 
One of the possible explanations for this phenomenon was suggested by  
\cite{baryshev93, baryshev98} 
in terms of gravitational lensing 
of remote quasars by intermediate-mass objects ($10^6 - 10^7$\,M$_\odot$) 
located in the vicinities of foreground galaxies. These could be globular clusters or satellite
dwarf galaxies. Gravitational lensing by intermediate-mass objects ({\it mesolensing}) 
differs from microlensing \citep{linder88} by much longer duration of the  
effect.
It is also different from lensing by large-mass foreground objects which usually
splits the lensed object image into multiple images, whereas  mesolensing does not 
split images and only results in brightness amplification.   

Then \cite{raikov16} suggested that the same mesolensing effect might be 
responsible for the brightness amplification of some superluminous 
quasars at high redshifts. The problem is that within the standard cosmological model 
some of them have luminosities $L_{\rm bol} > 10^{14}\,L_\odot$. 
By assuming that quasar energy is emitted via the mechanism of accretion 
onto a supermassive blackhole, the corresponding  mass of such a blackhole  
should exceed $10^{10}\,M_\odot$. On the other hand, the age of high-redshift objects 
(less than 0.7\,Gyr) is too small to explain their formation in the 
early Universe. 

In the last few years, the number of catalogued high-redshift quasars has  
dramatically increased \citep{flesch15, gattano18, ross20, flesch21}, and so has the number
of known superluminous quasars at high redhshifts.  In X-rays, the all-sky survey by 
the {\it Spektr-RG} (SRG) space observatory also produces a growing number 
of high-redshift quasars with X-ray luminosities exceeding $10^{46}$\, erg/s 
and masses of their blackholes exceeding $10^{9}\,M_\odot$ \citep{khorunzhev21}. 

Most of the high-redshift superluminous quasars are seen as single
sources, but we know \citep{baryshev97} that their brightnesses could be enhanced by mesolensing
without splitting source images, provided that the mass profile in the gravitational
lens is of the King-type \citep{king62}. There are numerous objects with such mass profiles
in the vicinities of foreground galaxies, so one would expect a large fraction
of quasars to be lensed. 

This fraction can be estimated by exploring statistical properties of 
available quasar catalogues,  the most convenient one
being the large astrometric catalogue of quasars LQAC-4 \citep{gattano18}
because it includes absolute magnitudes for different wavelength bands. 

\begin{figure}
\vspace{-0.9cm}
\hspace{-0.1cm}
\epsfig{figure=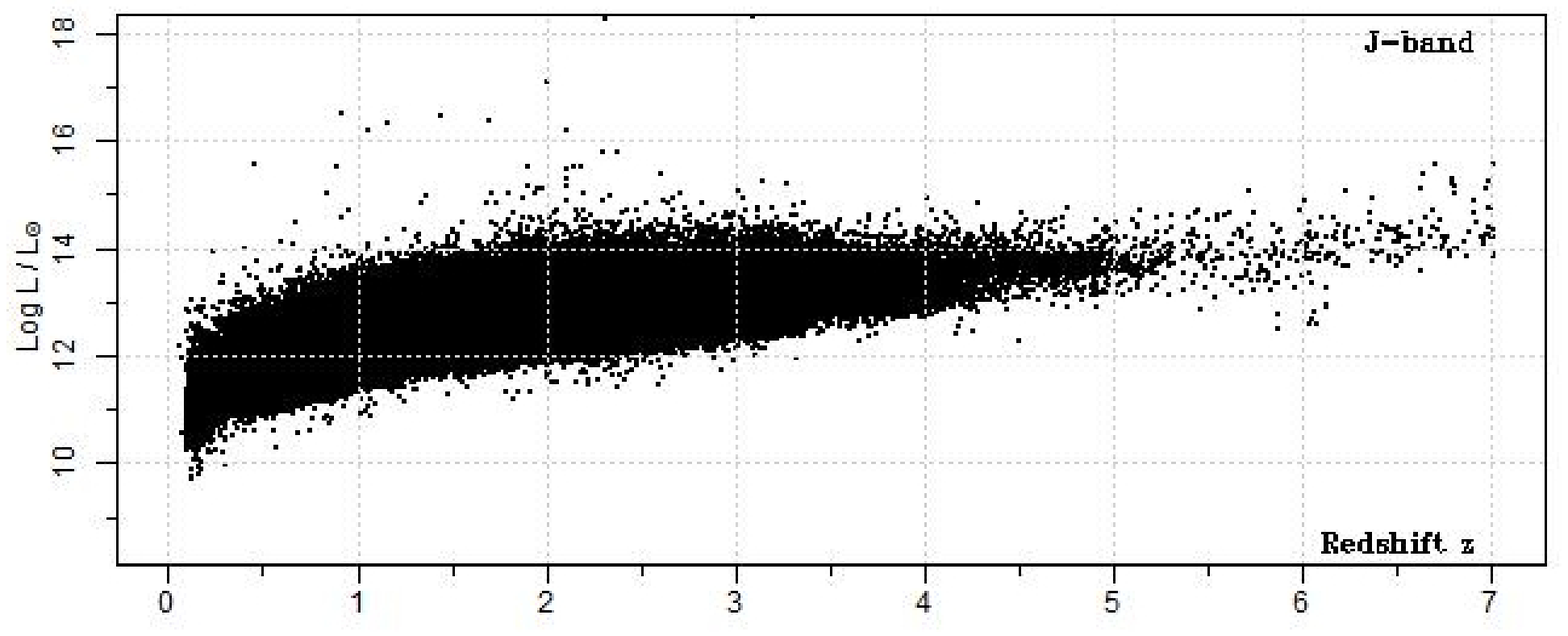,width=6.3cm}
\epsfig{figure=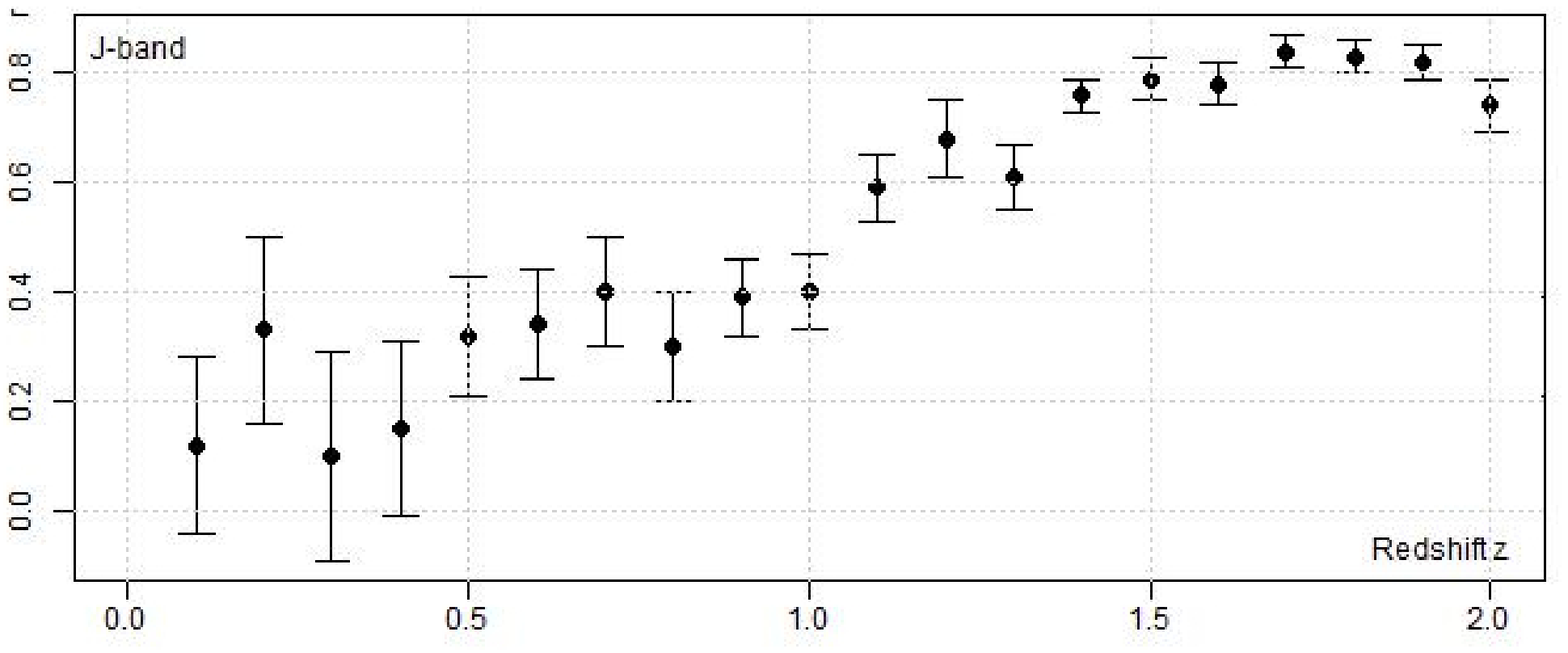,width=6.0cm}
\vspace{-0.4cm}
\caption{{\small {\it Left:} redshift distribution of quasar luminosities from the
LQAC-4 catalogue for the infrared band {\tt J};
{\it Right:}~fraction 
$r = (n_{\tt bright} - n_{\tt faint}) / (n_{\tt bright} + n_{\tt faint})$ indicating the 
excess of bright quasars over the faint ones in the luminosity histograms 
for redshift slices within $0.1 < z < 2$ and with $\Delta z=0.1$ for each slice
(two examples of these histograms are shown in Fig.\,\ref{fig:qso_zhist}
for $z=0.2$ and $z=0.8$).\vspace{-0.5cm}}} 
\label{fig:qso_bandj}
\end{figure}

We have converted these absolute magnitudes to bolometric 
luminosities by using the average spectral energy distribution
of quasars published by \cite{krawczyk13}.
The left panel of Fig.\,\ref{fig:qso_bandj} shows the redshift distribution of 
these luminosities based on the infrared band {\tt J} from LQAC-4. 

\noindent
The infrared band {\tt J} is likely to be less affected by the observational selection effect
than the other wavelength bands.    
However, even in this case, an observational
selection effect takes place for 
redshifts $z > 3$, because one can see  that for these redshifts 
the quasar luminosity-evolution stripe narrows at its low-luminosity fringe,
as shown in Fig.\,\ref{fig:qso_bandj} (left).
By contrast, the width of this stripe  
at $ 0.5 < z < 2.0$ remains approximately constant, which indicates that 
the observational selection effect for low-redshift quasars is minimal
or absent (indeed, one can expect that quasars, by being the brightest objects 
in the Universe, are all captured by modern telescopes at low redshifts). 

\begin{figure}
\vspace{-0.8cm}
\begin{tabular}{ll}
\hspace{-0.1cm}
\epsfig{figure=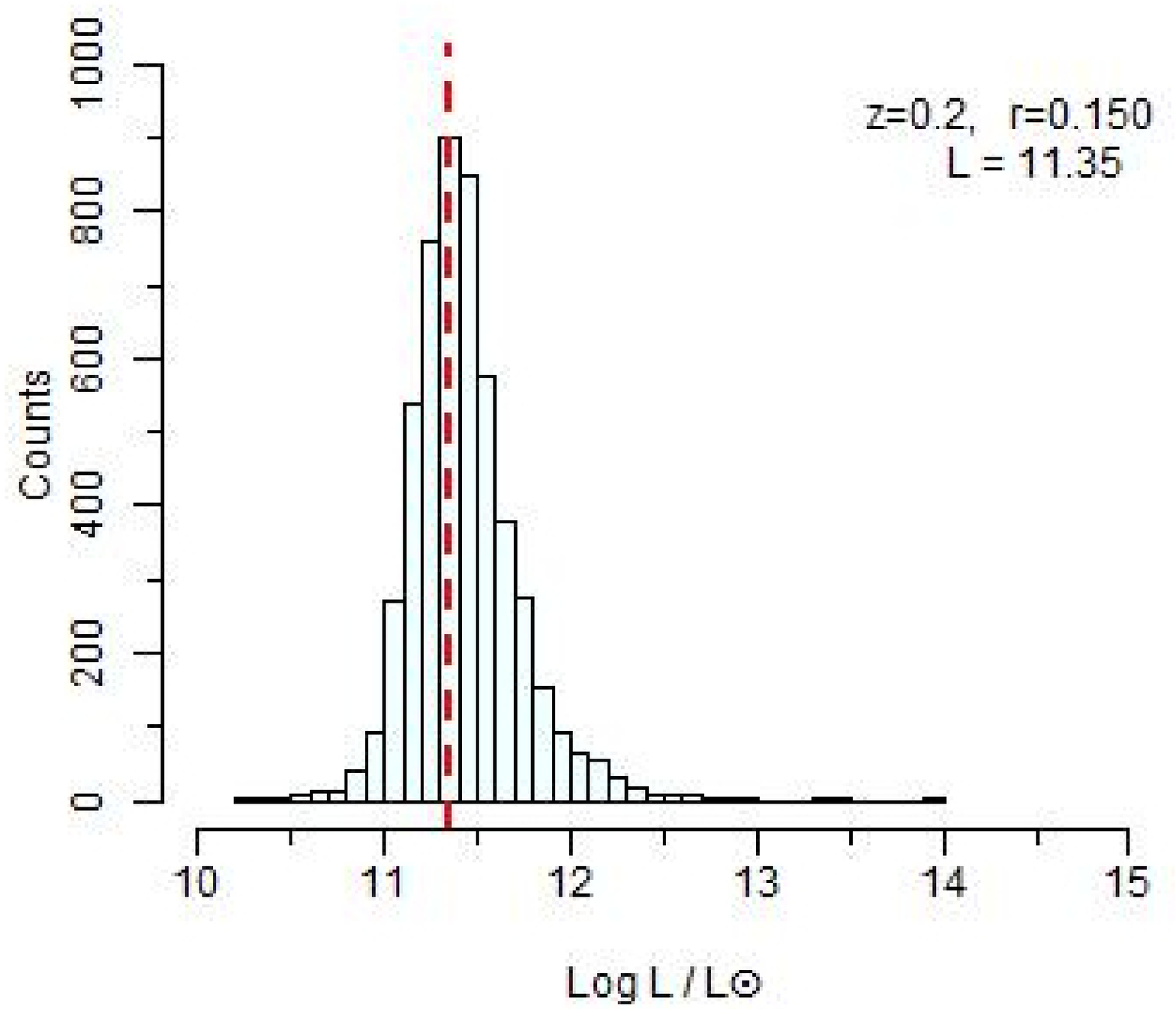,width=5cm} &
\epsfig{figure=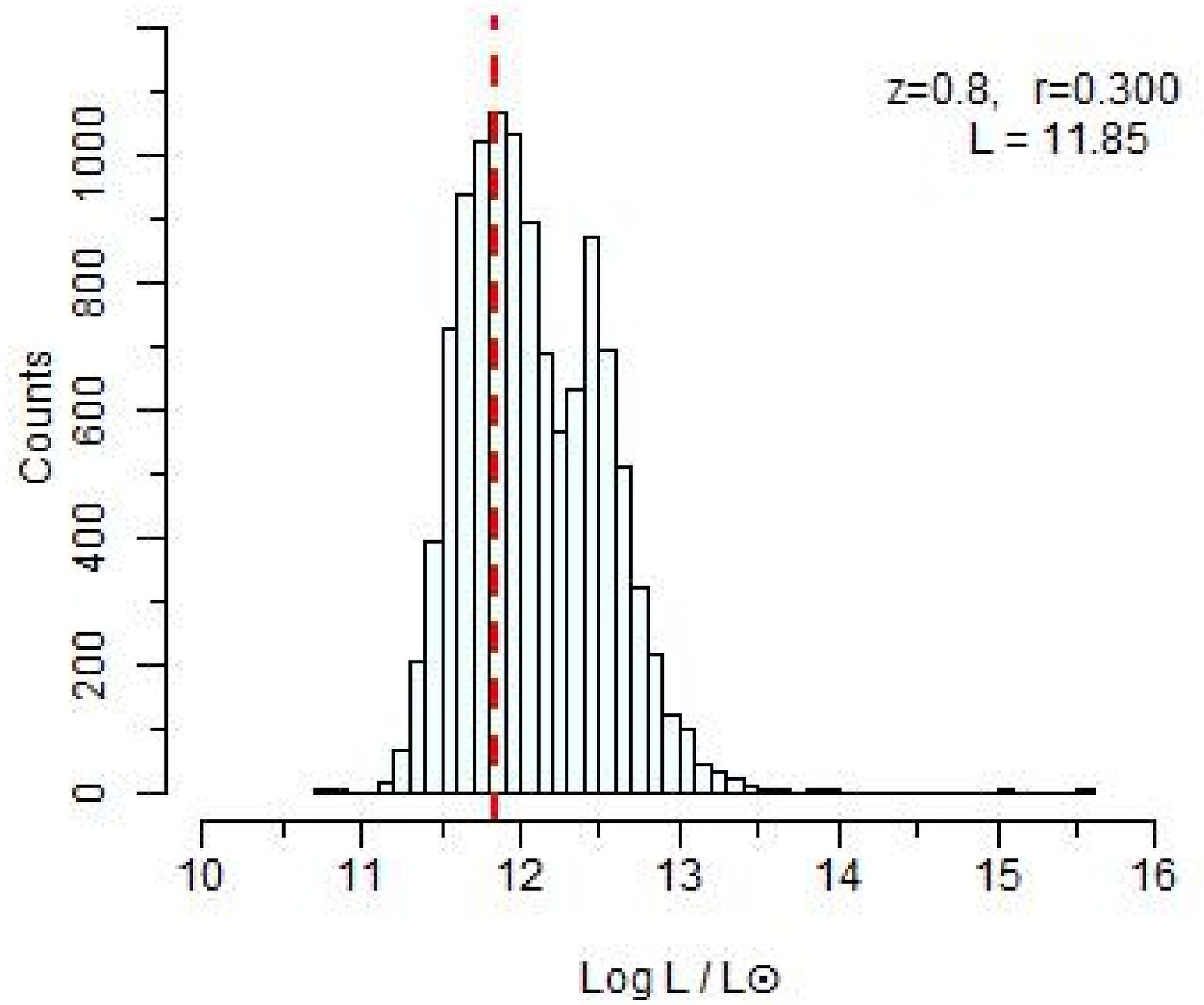,width=5.0cm} \\
\end{tabular}
\vspace{-0.3cm}
\caption{{\small ~Two examples of quasar luminosity histograms for redshift slices 
at $z=0.2$ and $z=0.8$, with the slice width $\Delta z=0.1$, used for calculating the histogram asymmetries 
$r=0.15$ and $r=0.30$, respectively. The positions of the luminosity peaks, $L$,
are indicated at the upper-right corners of each plot and also by the vertical dashed   
lines. \vspace{-0.3cm}  
}} 
\label{fig:qso_zhist}
\end{figure}
%

\noindent
We have built luminosity histograms for the quasar redshifts ranging from $z=0.1$ to $z=2.0$,
using a redshift interval 0.1 and the same width for each redshift slice $\Delta z=0.1$.
By counting the number of bright and faint objects, $n_{\tt bright}$ and $n_{\tt faint}$,
above and below the luminosity $L$ corresponding to the histogram maximum,
we can estimate the asymmetry -- the excess of the number of
bright quasars over the faint ones, 
$r=(n_{\tt bright} - n_{\tt faint}) / (n_{\tt bright} + n_{\tt faint})$,
which gives the fraction of gravitationally lensed quasars,
assuming that 
the natural luminosity distribution of quasars at each redshift slice 
must be symmetrical with respect to the
average luminosity.           

As we can see from the right panel of Fig.\,\ref{fig:qso_bandj},
the number of gravitationally lensed  quasars is growing with redshift, which 
is expected, as the number of foreground galaxies and their surrounding mesolenses
(globular clusters)
increases proportionally to the foreground volume. The negligibly small fraction of gravitationally
lensed quasars at small redshifts reaches 80\% at the redshift $z=1.5$.
Beyond this point, the linear growth of $r$ breaks down
and the fraction of lensed quasars becomes underestimated, which is likely to be  
caused by the observational selection effect. 

The calculations for $z>2$ give us 
an underestimated fraction $r \approx 30$\%. We can thus conclude (with caution) that 
the fraction of gravitationally lensed quasars is, at least, 30\%. However,
our calculations suggest a more radical conclusion that practically all of the high-redshift
quasars (say, for $z>5$) are gravitationally lensed, and that for some of them the geometrical 
configuration between the lensed quasars, the gravitational lens and the 
observer is such that the lensed quasars appear as 
superluminous objects, with their luminosities amplified by a factor of 
hundreds to thousands.

Therefore, the masses of supermassive blackholes associated with these quasars 
could actually be much smaller, which would resolve the conflict between the age
of the Universe corresponding to the redshifts of superluminous quasars
and the time needed for the formation of their supermassive blackholes.    

The time scales of the quasar brightness variability due to their gravitational lensing
on globular clusters can be from years to a few thousand years \citep{baryshev97},
depending on the source-to-lens geometrical configurations. Some  
smaller-time-scale variability can be expected due to microlensing on individual stars
belonging to globular clusters. 

As an example, the superluminous X-ray quasar
SRGE J170245.3+130104 at redshift $z=5.5$ has reportedly reduced its brightness
by half between the first and second all-sky surveys by the SRG space observatory 
\citep{khorunzhev21}. This might be caused by either internal change in the quasar 
accretion process or by the change in the quasar-to-lens geometrical configuration.
In the latter case, one would expect some further reduction of its observed brightness 
during the next forthcoming all-sky surveys by the SRG. 

Foreground objects acting as mesolenses are unlikely to be observed directly or 
spectrally by having luminosities much smaller than those of quasars. However,
since the lensing effect is the same for all wavelengths, 
comparing light curves in different spectral bands could be regarded as an
observational test for mesolensing.
    
\noindent 
{\small   
{\textbf{Acknowledgement.} {\it This research has been partly supported by the St. Petersburg State University's project \#~73555239.}
}

%
%
\bibliographystyle{plainnat}
{\let\clearpage\relax

}
\end{document}